\documentclass[prl,showpacs,preprintnumbers,preprint]{revtex4}

\usepackage{graphicx}

\begin{document}

\title{Synchronization in semiconductor laser rings}

\author
{Javier M. Buld\'{u}$^{1}$, M. C. Torrent$^{2}$ and Jordi Garc\'{\i}a-Ojalvo$^{2}$}

\affiliation{$^{1}$Nonlinear Dynamics and Chaos Group, Departamento de Ciencias de la Naturaleza y F\'{\i}sica Aplicada,
Universidad Rey Juan Carlos, E-28933 M\'ostoles, Spain\\
$^{2}$Departament de F\'{\i}sica i Enginyeria Nuclear, Universitat Polit\`ecnica de Catalunya,Colom 11, E-08222 Terrassa, Spain\\}

\pacs{42.65.Sf, 05.45.Xt, 42.55.Px}
\date{\today}

\begin{abstract}
We examine the dynamics of semiconductor lasers coupled in a ring configuration. The lasers, which have stable output intensity when isolated, behave chaotically when coupled unidirectionally in a closed chain. In
this way, we show that neither feedback nor bidirectional coupling is necessary to induce chaotic dynamics at
the laser output. We study the synchronization phenomena arising in this particular coupling architecture, and discuss its possible application
to chaos-based communications. Next, we extend the study to bidirectional coupling and propose an 
appropriate technique to optical chaos encryption/decryption in closed chains of mutually coupled semiconductor lasers.
\end{abstract}

\maketitle


\section{Introduction}

Since the seminal work of Cuomo and Oppenheim \cite{cuo} demonstrating the potential of chaotic systems for 
information encoding, communications through chaotic carriers have been
increasingly studied \cite{koc,xia,par,uch}. Despite being first implemented in electronic circuits, the technological importance of optical communications
have recommended the use of chaotic lasers as emitters and receivers in chaos-based
communication systems \cite{col,mir,van}. Particularly, semiconductor lasers, which are ubiquitously used in optical communications, have
been widely studied \cite{kra}. Due to their fast dynamics, they have been
used to transmit information with bit-rates on the order of GHz \cite{san,pau}.
More recently, a field experiment using commercial optical networks has demonstrated
the ability and robustness of chaotic semiconductor lasers to transmit high-bit-rate messages in real-world 
conditions \cite{arg}.
Both single-mode \cite{kra} and multimode \cite{mol,bul} lasers have proved
adequate for optical chaotic communications (in the latter, with the possibility of multiplexing different messages through the 
longitudinal modes of the laser).

A solitary semiconductor laser is not able to have chaotic dynamics since
it is a class B laser \cite{agr,wei} (i.e. a two-dimensional dynamical system). To that end, some kind of external perturbation must be
considered. In the case of all-optical systems, external optical feedback has been the paradigmatic
way of induce chaos in semiconductor lasers \cite{lan}. Alternative ways of inducing chaotic dynamics in semiconductor lasers have
been via injection from another chaotic laser \cite{mir}
or by mutual injection between otherwise stable lasers \cite{hei,mul}.
In the present paper we propose a different technique to induce chaotic oscillations on coupled semiconductor lasers. 
By coupling isolated semiconductor lasers in closed loop
chains we obtain chaotic behavior even in the absence of optical feedback or bidirectional coupling. We investigate
the synchronization of the destabilized lasers and observe that, despite the high correlation between their outputs,
the lasers have transient states of desynchronization that prevent this configuration from constituting a robust technique
for chaotic encryption. Next, we extend the study to bidirectional coupling in the same kind of 
coupled chains. In this case, we observe identical synchronization between lasers, which 
synchronize with zero-lag despite the delay in the transmission channel. When a message is introduced by modulating the pumping
current of any of the lasers, it is possible to recover the message at any of its neighbors. In this way, we propose this straightforward configuration as a method of 
message encryption within a restricted community of users.


\section{The ring configuration}

\begin{figure}[!htb]
\centerline{
\includegraphics[width=12cm,clip]{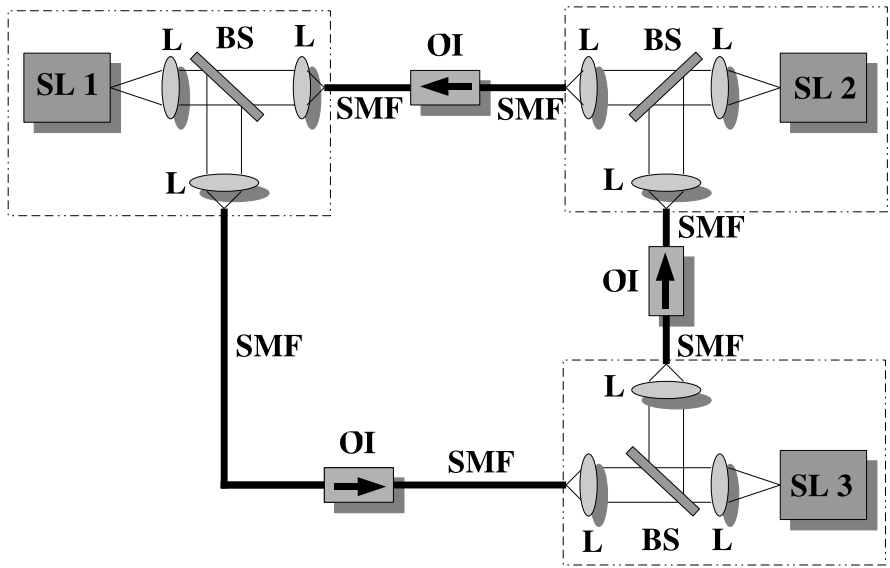}
} \caption[sdh]{
Schematic setup of the closed loop configuration. Three semiconductor lasers (SL1, SL2 and SL3)
are unidirectionally coupled thanks two optical isolators (OI) placed in the single mode fiber (SMF).
A lens (L) collimates the laser beam and a beam-splitter (BS) splits its path. 
} 
\label{fig:fig1}
\end{figure}

Figure \ref{fig:fig1} shows a schematic representation of the system proposed here.
The output beam of each laser is split in two paths with a $50\%$ beam-splitter (BS). In this way each laser
is able to receive light from a neighbor and, at the same time, send part of 
the output to the other neighbor. Once divided, the laser
beam is collimated into a single mode optical fiber (SMF), which acts 
as the transmission channel. Optical isolators (OI) guarantee unidirectional coupling between the lasers \footnote{Note that the setup could be further 
simplified if lasers with 
two open facets were considered.}.

Equations describing the dynamics of the configuration explained above can be easily obtained
from those of a unidirectionally injected semiconductor laser \cite{tar}. Two variables describe
the dynamics of each laser, the carrier number $N(t)$ and the slowly varying envelope of the
complex field $E(t)$: 
\begin{eqnarray}
\dot{E}_m(t) & = & \frac{1}{2} (1+i \alpha) \left[ G_m - \gamma \right]
E_m(t)\nonumber\\
& + & \kappa_{in} e^{-i \omega_0 \tau_c} E_{in}(t-\tau_c)
+\sqrt{2\beta N}\xi (t)\label{eq:field}\\
\dot{N}_m(t) & = & \frac{I_m}{e} - \gamma_e N_m - G_m |E_m|^2 \label{eq:carrier}
\end{eqnarray}
where the subindex $m$ indicates the laser number (m=1,2,3) and the subindex $in$ corresponds
to the laser whose light is injected into laser $m$. Parameters of Eqs.~(\ref{eq:field})-(\ref{eq:carrier})
are the linewidth enhancement factor $\alpha$, the photon decay rate $\gamma$, the
strength $\kappa_{in}$ of injection (unidirectional coupling), the coupling time $\tau_c$, 
the free-running frequency of the laser $\omega_0$,
the pumping current $I$, the elementary charge $e$ and the carrier decay rate $\gamma_{e}$.
The last term of Eq.~(\ref{eq:field}) accounts for spontaneous emission, $\xi(t)$ being a Gaussian white 
noise term of zero mean and unity intensity, and $\beta$ measuring the noise strength.
The material gain $G_m$ is given by:
\begin{eqnarray}
G_m = \frac{g(N_m - N_0)}{1+\epsilon |E_m|^2},
\end{eqnarray}
where $g$ is the differential gain, $N_0$ is the transparency inversion and $\epsilon$ is the saturation coefficient. 
The term $\omega_0 \tau_c$ [see second term of Eq. (\ref{eq:field})], known as the injection phase, is set to zero for simplicity;
numerical simulations have shown that setting $\omega_0 \tau \neq 0$ does not change the results that follow. 
The injection phase can be controlled experimentally by fine tuning of either the lasing frequency or the cavity length. 
Table \ref{tab:tab01} summarizes the values of the parameters described above, which are assumed equal for all lasers.
Numerical simulations with slight ($<3\%$) parameter mismatch show
results similar to those obtained with equal parameters.

{\setlength\arraycolsep{-10pt}
\begin{table}[htb]
\begin{center}
\begin{tabular}{@{}lcl@{}}
Description&  &Value\\
\hline
\hline
Linewidth enhancement
factor & $\alpha$ & $4.0$\\
Cavity decay rate & $\gamma$ & $0.48$~ps$^{-1}$\\
Carrier decay rate & $\gamma_e$ & $6.89\times 10^{-4}$~ps$^{-1}$\\
Spontaneous emission noise & $\beta$ & $0.1\times 10^{-9}$~ps$^{-1}$\\
Injection current & I & $1.010\times I_{th}$\\
Saturation coefficient & $\epsilon$ & $0.0$\\
Differential gain & $g$ & $1.2\times 10^{-8}$~ps$^{-1}$\\
Transparency inversion & $N_0$ & $1.25\times 10^{8}$\\
Injection strength & $\kappa_i$ & $0.020$~ps$^{-1}$\\
Coupling time & $\tau_c$ & $2.0$~ns\\
Injection phase & $\omega_0 \tau_c$ & 0 \\
\hline
\end{tabular}
\caption{\label{tab:tab01}
Parameter values (the same for all lasers) used in the simulations.
}
\end{center}
\end{table}
}


\section{Chaotic dynamics and synchronization}

As mentioned in the introduction, feedback has been the most usual way to 
induce chaotic dynamics in a semiconductor laser. More recently, mutual coupling has been
shown to be an alternative technique to induce aperiodic dynamics. For the case of a solitary laser
unidirectionally injected into a second one, and in the absence of optical detuning, chaotic
dynamics do not arise, at least for low to moderate injection strengths $\kappa_c$. Furthermore,
the coupling time $\tau_c$ does not have any relevance on the local dynamics of the lasers. Under 
these conditions, the {\em receiver} laser is injection locked, and adjusts its phase to that
of the {\em transmitter}, with both lasers having constant power. When more lasers
are introduced in the unidirectional chain, phase locking and constant power hold throughout the
chain. Nevertheless, there is a straightforward way of destabilizing the whole array of lasers: closing the chain.

\begin{figure}[!htb]
\centerline{
\includegraphics[width=12cm,clip]{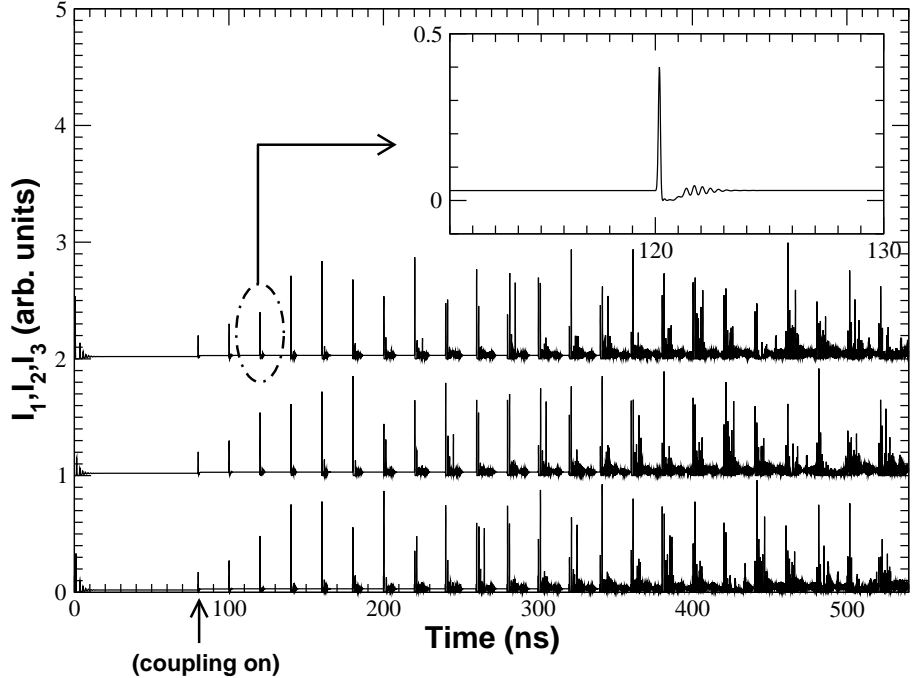}
} \caption[sdh]{
From stable to chaotic dynamics: Output intensity (vertically shifted) of three semiconductor lasers SL1, 
SL2 and SL3 unidirectionally coupled in a ring configuration. Injection is initially blocked 
until lasers achieve CW operation. At $t_{on}=80$~ns,
unidirectional injection is allowed and perturbations are transmitted through the chain.
The inset shows how the perturbation is followed by the relaxation oscillations at the output
intensity. After a certain transient, the dynamics of all lasers become aperiodic.
} 
\label{fig:fig2}
\end{figure}

If a perturbation is applied to the first laser of an open chain, it is transmitted
to the following laser, which in turn sends the perturbation to the following node of the chain, and so on.
When the perturbation goes through each of the lasers in the chain, the laser returns to its stable output via
characteristic relaxation oscillations \cite{agr}. On the other hand, if the chain is closed, any 
perturbation introduced into the system will remain on it. Feedback of a traveling perturbation onto itself, upon interaction with the relaxation oscillations excited by it during the previous round-trip,
fully destabilizes the 
laser intensities after a certain number of laps within the closed chain, leading to
chaos in all the lasers in the ring. Figure~\ref{fig:fig2} shows an example of the transition
to chaos for the configuration depicted in Fig.~\ref{fig:fig1}. In this particular case, we have neglected
spontaneous emission in order to observe the propagation of the perturbations. The coupling time
is set to $\tau_c=2$~ns for all pairs of lasers. Under these conditions, we block the coupling, turn on the lasers
and wait for a short transient 
until they achieve their constant output. When all lasers have constant power, we unblock the coupling
($t_{on}=80$~ns) and a sudden jump, due to the injection, is observed in the outputs of all lasers. The perturbation
disappears after some oscillations of the laser output, as can be observed at the inset of Fig. \ref{fig:fig2}.    
However, the intensity perturbation is sent to the following laser and returns after
a complete loop, interacting with the perturbation remaining from the previous
roundtrip. Each roundtrip, the perturbation is further stretched in time and, after 
a certain number of roundtrips (see temporal evolution in Fig.~\ref{fig:fig2}), it leads
to chaotic dynamics in all three lasers.

\begin{figure}[!htb]
\centerline{
\includegraphics[width=12cm,clip]{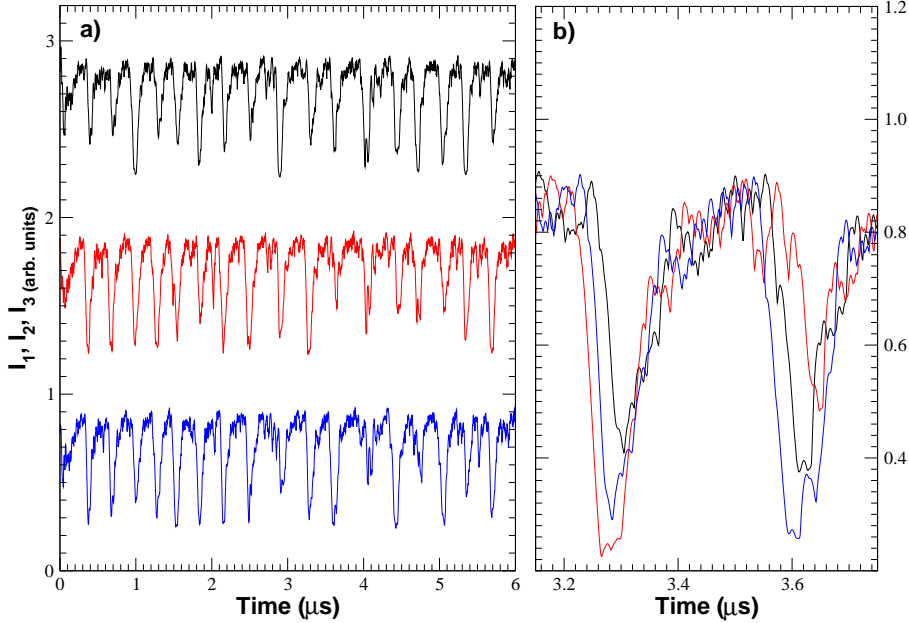}
} \caption[sdh]{
In (a): Laser intensities of SL1 (upper), SL2 (middle) and SL3 (bottom), which have been vertically shifted
in order to ease comparison. All intensities have been filtered by a second-order Butterworth
filter, in order to simulate the unavoidable filtering of the photodetectors in a real experiment.
In (b): A zoom shows that despite all laser intensities fall in chain, the role
of the first laser is exchanged.
} 
\label{fig:fig3}
\end{figure}

Once the lasers have reached the unstable behavior, we can analyze the relationship between their outputs.
Figure~\ref{fig:fig3} shows the temporal evolution of the laser outputs, once filtered by 
a low pass filter. In this way, we simulate the filtering effects of a photodetector in a 
real experiment \cite{suk}, thereby unveiling the low-frequency fluctuation
dynamics \cite{san94} of the system, consisting on intensity drops at frequencies
much lower than those of the fast intensity pulses.

Figure \ref{fig:fig3}(a) shows that despite synchronization is not perfect, all laser intensities
fall together. A closer look at the time series [Fig. \ref{fig:fig3}(b)]
gives away some additional information. First, intensity dropouts occur sequentially, with 
a time delay between the lasers equal to the coupling time $\tau_c$. Second, the laser
that falls first, inducing enchained dropouts, is not always the same.
This behavior is typically associated with two bidirectionally coupled semiconductor lasers,
which synchronize with a time delay of $\tau_c$, the role of leader switching randomly from one laser to the other, provided the two lasers have equal frequencies
\cite{hei}.

Let us evaluate quantitatively the quality of synchronization. With this aim we compute
the cross-correlation function $C(\Delta t)$ between pairs of unfiltered intensity time series,
\begin{equation}
C_{ij}(\Delta t)=\frac{\langle(I_i(t)-\langle I_i \rangle)(I_j(t+\Delta t)-\langle I_j\rangle)\rangle}
{\sqrt{\langle(I_i(t)-\langle I_i\rangle)^2\rangle\langle(I_j(t)-\langle I_j\rangle)^2\rangle}}\,,
\label{eq:correlation}
\nonumber
\\
\end{equation}
where $I_i$ and $I_j$ represent the output intensity of lasers $i$ and $j$, respectively ($i,j=1,2,3$) and the brackets indicate time averaging.
In this way we compute the correlation between time series for different shifts in the time-axis,
obtaining the quality of the synchronization ($-1<C(\Delta t)<1$) and the delay between series.

\begin{figure}[!htb]
\centerline{
\includegraphics[width=12cm,clip]{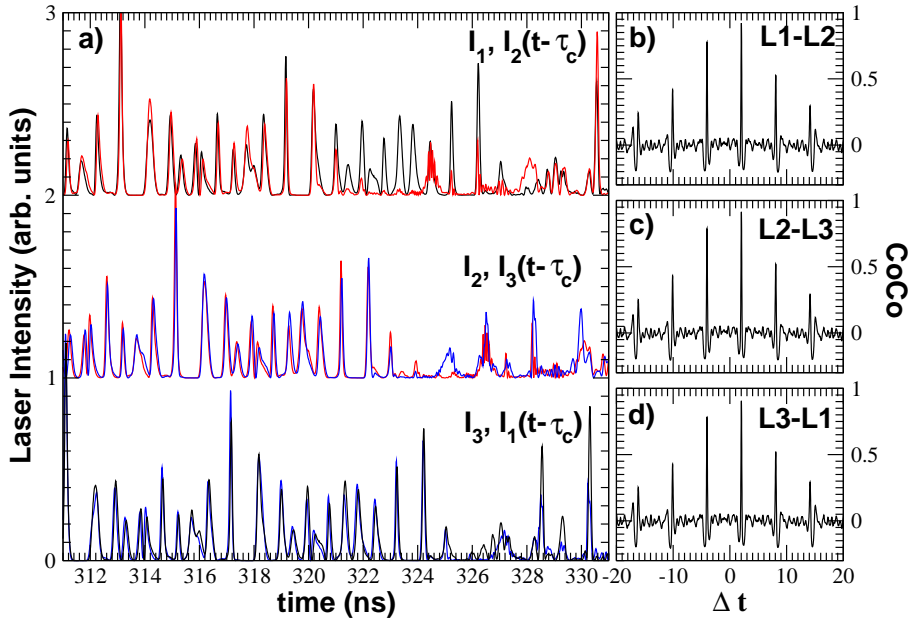}
} \caption[sdh]{
Output intensities of SL1, SL2 and SL3 (a), in pairs, and the corresponding cross-correlation
function (b). In (a) one of the time series shifted $\tau_c=2$~ns in order to show time-lagged synchronization 
between lasers. In (b) maxima are placed at $\Delta t_{\rm max}=2$~ns and have a value of (from top to bottom):
$C_{L1-L2}(\Delta t_{\rm max})=0.92$,
$C_{L2-L3}(\Delta t_{\rm max})=0.91$ and
$C_{L3-L1}(\Delta t_{\rm max})=0.91$. 
} 
\label{fig:fig4}
\end{figure}

In Fig.~\ref{fig:fig4} we plot the three unfiltered laser outputs, together with
the cross-correlation function $C(\Delta t)$. Time series are plotted in pairs, with one of the laser
outputs shifted a time $\tau_c$, in order to reveal synchronization. We can observe that
regions of synchronized dynamics coexist with others where synchronization is lost. This
point is not reflected when computing the cross-correlation function between lasers, which
has maxima, in all cases, with values higher than $0.9$. This fact reveals that synchronization
between lasers is good and that the unsynchronized episodes are much shorter than the synchronization
regions. Furthermore, we can observe how the cross-correlation peaks at $\Delta t=2$~ns, indicating a time delay
between lasers equal to the coupling time $\tau_c=2$~ns.

Summarizing the main observations concerning the synchronization between the three
undirectionally coupled chaotic lasers, we can say that: a) they show good synchronization, b) they exchange the leader-laggard role randomly in time, and
c) they exhibit short episodes of unsynchronized
behavior. The last two points are crucial to evaluate the ability of this configuration to
encrypt/decrypt a message. Encryption with chaotic carriers relies in the synchronization between
two chaotic systems, if synchronization is not maintained it is not possible to recover
the encrypted message. 
In this sense, the cross-correlation function, which is often used as a way of measuring the quality of synchronization,
does not work in evaluating the ability of the system to recover an encrypted message. Since the cross correlation is only
a mean value of two time series, we do not know if deviations from $C(\Delta t)=1$ (perfect synchronization)
are due to an offset error, which could be filtered, or to short transients where synchronization
is completely lost. In order to distinguish between both cases we compute the sliding cross-correlation,
which consists on the maximum of the cross-correlation function evaluated in short temporal windows for a given
shift of the time series (in this case $\Delta t=2$~ns). In this way we obtain the instantaneous correlation
(within a short window) between laser outputs.

\begin{figure}[!htb]
\centerline{
\includegraphics[width=12cm,clip]{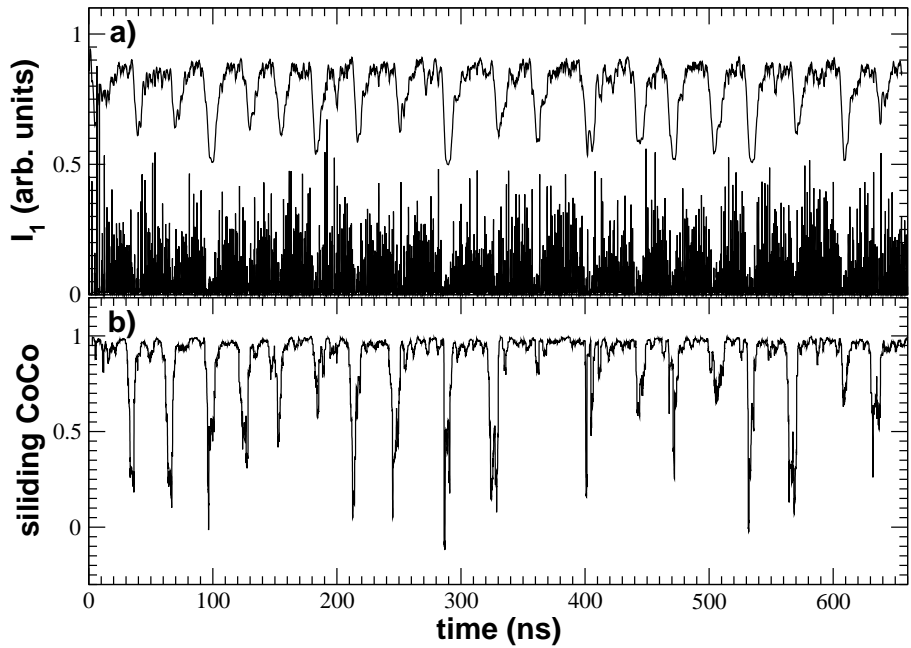}
} \caption[sdh]{
(a) Filtered (upper trace) and unfiltered (lower trace) output intensities of SL1. The
former has been
vertically shifted in order to ease comparison. (b) Sliding cross-correlation between
the output intensities of SL1 and SL2.
} 
\label{fig:fig5}
\end{figure}

Figure \ref{fig:fig5} shows the temporal evolution of SL1 intensity $I_1(t)$ (a) and the corresponding 
sliding cross-correlation (b) with SL2 intensity $I_2(t)$.
We have plotted both the unfiltered and filtered
intensity of SL1 for better reference. We can see how, despite values of the cross-correlation are close
to 1, there are strong decays which correspond with transients where synchronization is lost. This fact
indicates that closed rings of unidirectionally coupled semiconductor lasers are not adequate to recover 
an encrypted message, since this requires not only high cross-correlation between output intensities, but also continuous synchronization.
It is worth noting that episodes of unsynchronized behavior are related with the intensity dropouts
of the lasers, as can be observed by comparing the filtered output intensity [Fig.~\ref{fig:fig5}(a)] with the sliding
cross correlation [Fig.~\ref{fig:fig5}(b)]. As shown in \cite{bul06}, it is during the intensity turn-off that the lasers
are prone to lose synchronization, since their optical frequency shifts drastically.


\section{Bidirectional communication}

\begin{figure}[!htb]
\centerline{
\includegraphics[width=12cm,clip]{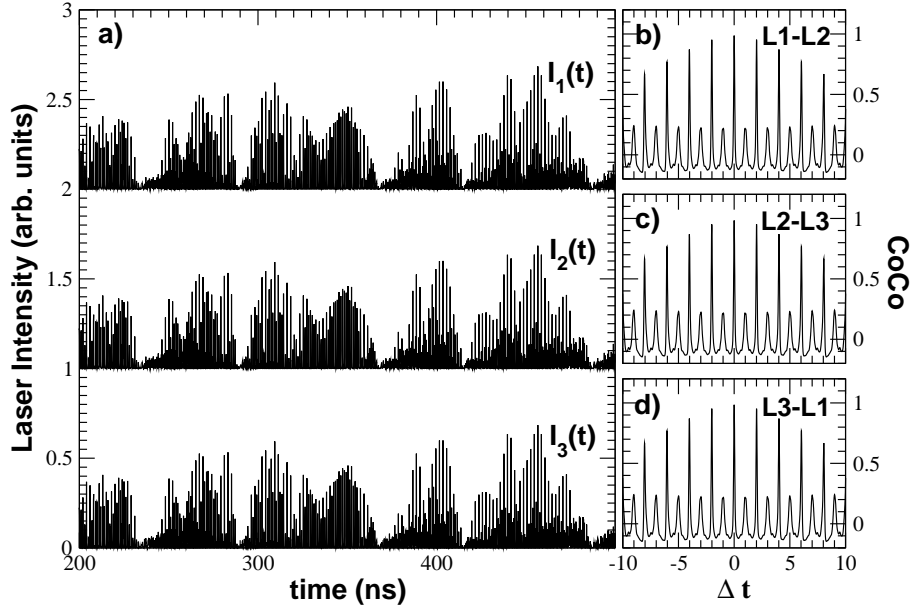}
} \caption[sdh]{
Output intensities of SL1, SL2 and SL3 (a), and the corresponding cross-correlation
function (b). In (b), maxima are placed at $\Delta t=0$~ns, which indicates that synchronization occurs at zero lag.
Value of the maxima are (from top to bottom):
$C_{L1-L2}(\Delta t_{\rm max})=0.986$,
$C_{L2-L3}(\Delta t_{\rm max})=0.983$, and
$C_{L3-L1}(\Delta t_{\rm max})=0.985$. 
} 
\label{fig:fig6}
\end{figure}

The previous results seem to indicate that a ring configuration would not be a suitable way of transmitting information within a community of users: despite unidirectional injection leads to chaotic synchronized
behavior, it does not guarantee permanent synchronization. In fact, similar behavior is reported
in the case of two bidirectionally coupled SL, where again synchronization does not help in recovering the encoded information \cite{mul}.
Nevertheless, we can combine both ingredients, closed-loop chain and bidirectional coupling, and observe
what kind of synchronization is obtained. As shown in \cite{fis06}, it is possible to synchronize
a linear chain of
three mutually coupled semiconductor lasers, so let us go one step further by coupling the 
lasers in a bidirectional ring. Furthermore, it is a simpler configuration, since 
the setup would be the one of Fig. \ref{fig:fig1}
where the optical isolators (OI) have been removed.
Figure~\ref{fig:fig6} shows the output intensities of the three lasers with their corresponding
cross-correlation function. The time series gives some insights about what we are going to find in the
cross-correlation plot: the three laser intensities are surprisingly similar. Lasers are synchronized both at
low and high time scales and synchronization holds continuously. This point is reflected at
the cross-correlation, which for all pairs of lasers has a maximum higher than $0.98$,
corresponding to a very good level of
synchronization. It is worth noting that the best correlation is obtained at $\Delta t=0$, indicating
that the three lasers synchronize isochronally, irrespective of the coupling time between them.
This phenomenon, known as zero-lag synchronization, has been recently observed 
between the outer lasers of a linear chain \cite{fis06}. 
Finally, when computing the sliding cross-correlation between
pairs of lasers (not shown here), we obtain values close to unity which hold through time, a fact
that suggests that we have found a suitable configuration for message encryption/decryption.

The next step is to introduce and encode a message at one of the lasers and try to recover it at another one.
We are going to use a standard encryption method known as Chaos Shift Keying (CSK) \cite{uch}.
This technique is based in the chaos-pass filtering properties of the chaotic systems. When 
two chaotic systems synchronize, they filter out small perturbations from the
coupling signal, synchronizing
only with its chaotic part. The two chaotic systems need to be similar,
and mismatches in any of their internal parameters can induce
a loss of synchronization. In our particular case,  
the message is introduced at one of the lasers by varying slightly the pumping current of SL1.
In this way, the receiver laser will only synchronize when the pump currents of both transmitter and receiver
are equal. Thus, we can recover the message at SL2 or SL3 by computing the difference between the injection 
coming from SL1 and the output of the receiver laser. When the pump currents are different the lasers
will lose synchronization.
This procedure is shown at Fig.~\ref{fig:fig6}. In (a) we show the message that will be
introduced into the laser by pumping modulation. The amplitude of the message is $0.2\%$ of the pumping current
and must be as low as possible, in order to assure that the message is hidden at the chaotic output. Figure~\ref{fig:fig6}(b)
shows the output intensity of SL1, which completely hides the input message. The corresponding
spectrum of the signal (not shown here) is also covered by the spectrum of the chaotic carrier.
Figure~\ref{fig:fig6}(c) shows the difference between the input signal coming from SL1 and the output
of SL2. We can see episodes of unsynchronized dynamics that correspond to the modulation
of the pump current of SL1 (i.e. the message). Finally, by filtering and reshaping the subtracted signal we are
able to recover the encoded message.  

\begin{figure}[!htb]
\centerline{
\includegraphics[width=12cm,clip]{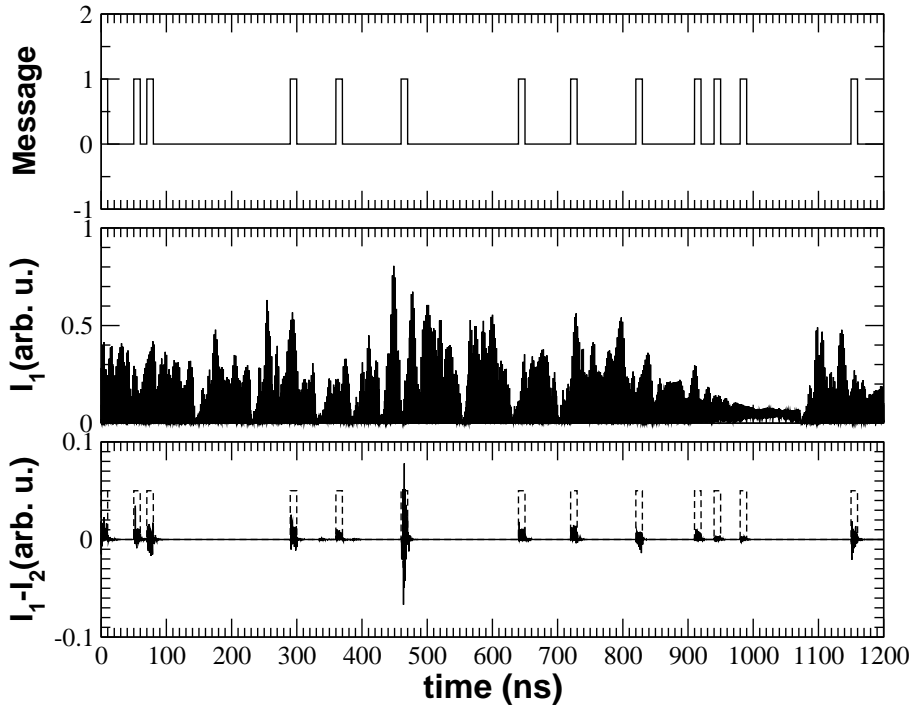}
} \caption[sdh]{
Example of message encryption/decryption: (a) message sent from SL1, (b) output intensity of SL1,
showing the message encryption and (c) intensity difference between SL1 and SL2 and recovered message 
(dashed line) after filtering and reshaping. 
} 
\label{fig:fig7}
\end{figure}


\section{Conclusions}

We have studied the phenomena of chaos synchronization and communication in rings of coupled 
semiconductor lasers. Unidirectional coupling is shown to induce
chaotic dynamics in the otherwise stable lasers. However, even though
a good degree of synchronization exists between the lasers in this case,
the existence of periods of synchronization loss, and the appearance of a leader-laggard
dynamics, does not render this configuration suitable for chaos-based communications.
The case of bidirectionally
coupled lasers, on the other hand, leads to zero-lag synchronization between all three lasers at all times, without periods of synchronization loss, and thus this configuration is more adequate for chaotic communications.
We believe that this technique could be applied to secure communications within a restricted community
of users. Furthermore, the experimental implementation is simpler than previous setups
where optical feedback at the transmitter lasers is required. 

Research supported by the Ministerio de Educaci\'on y Ciencia (Spain) and by the Generalitat de Catalunya. We thank Claudio Mirasso,
Ingo Fischer and Raul Vicente for fruitful discussions.


\begin{thebibliography}{18}

\bibitem{cuo} K.M. Cuomo and A.V. Oppenheim
``Circuit implementation of synchronized chaos with applications to communications",
{\em Phys. Rev. Lett.} {\bf 71}, 65--68 (1993).

\bibitem{koc} L. Kocarev and U. Parlitz,
``General Approach for Chaotic Synchronization with Applications to Communication",
{\em Phys. Rev. Lett.} {\bf 74}, 5028 (1995).

\bibitem{xia} J.H. Xiao, G. Hu and Z. Qu,
``Synchronization of Spatiotemporal Chaos and Its Application to Multichannel Spread-Spectrum Communication",
{\em Phys. Rev. Lett.} {\bf 77}, 4162 (1995).

\bibitem{par}  U. Parlitz, L. Kocarev, T. Stojanovski and H. Preckel
``Encoding messages using chaotic synchronization",
{\em Phys. Rev. E} {\bf 53}, 4351--4361 (1996).

\bibitem{uch} A. Uchida, F. Rogister, J. Garcia-Ojalvo and R. Roy,
``Synchronization and communication with chaotic laser systems",
{\em Progress in Optics} {\bf 48}, 203--341 (2005).

\bibitem{col} P. Colet and R. Roy,
``Digital communications with synchronized chaotic lasers",
{\em  Opt. Lett.} {\bf 19}, 2056--2058 (1994).

\bibitem{mir} C.R. Mirasso, P. Colet and P. Garc\'{\i}a-Fern\'andez,
``Synchronization of chaotic semiconductor lasers: application to encoded communication",
{\em  Phot. Tech. Lett.} {\bf 8}, 299--301 (1996).

\bibitem{van} G.D. Van Wiggeren and R. Roy,
``Communication with chaotic lasers",
{\em  Science} {\bf 279}, 2056--2058 (1999).

\bibitem{ahl} V. Ahlers, U. Parlitz and W. Lauterborn,
``Hyperchaotic dynamics and synchronization of external-cavity semiconductor lasers",
{\em  Phys. Rev. E} {\bf 58}, 7208--7213 (1998).

\bibitem{kra} {\em Fundamental Issues of Nonlinear Laser Dynamics}, Ed. B. Krauskopf
and D. Lenstra, AIP Conference Proceedings, New York (2000).

\bibitem{san} A. S\'anchez-D\'iaz, C.R. Mirasso, P. Colet and P. Garc\'{\i}a-Fern\'andez,
``Encoded Gbit/s digital communications with synchronized chaotic semiconductor lasers",
{\em  IEEE J. Quantum Electron.} {\bf 35}, 292--297 (1999).

\bibitem{pau} J. Paul, S. Sivaprakasam, P.S. Spencer, P. Rees and K.A. Shore,
``GHz bandwidth message transmission using chaotic diode lasers", 
{\em Electron. Letts}, {\bf 38}, 28--29 (2002).

\bibitem{arg} A. Argyris, D. Syvridis, L. Larger, V. Annovazzi-Lodi, P. Colet,
I. Fischer, J. Garc\'{\i}a-Ojalvo, C.R. Mirasso, L. Pesquera and K.A. Shore
``Chaos--based communications at high bit rates using commercial fibre--optic links",
{\em Nature} {\bf 438}, 343--346 (2005).

\bibitem{mol} J.K. White and J.V. Moloney,
``Multichannel communication using an infinite dimensional spatiotemporal chaotic system",
{\em  Phys. Rev. A} {\bf 59}, 2422--2426 (1999).

\bibitem{bul} J.M. Buld\'u, J. Garc\'{\i}a-Ojalvo and M. C. Torrent, 
``Multimode synchronization and communication using unidirectionally coupled semiconductor lasers", 
{\em IEEE J. Quantum Electron.} {\bf 4}, 640 (2004).

\bibitem{agr} G.P. Agrawal and N.K. Dutta,
``Semiconductor Lasers",
Kluwer Academic Press, Norwell (1993).

\bibitem{wei} C.O. Weiss and R. Vilaseca,
``Dynamics of Lasers",
VCH Publishing, Weinheim (1991).

\bibitem{lan} R. Lang and K. Kobayashi,
``External optical feedback effects on semiconductor injection laser properties",
{\em IEEE J. Quantum Electron.} {\bf 16}, 347 (1980).

\bibitem{hei} T. Heil, I. Fischer, W. Els{\"a}sser, J. Mulet and C.R. Mirasso,
``Chaos Synchronization and Spontaneous Symmetry-Breaking in Symmetrically Delay-Coupled Semiconductor Lasers",
{\em Phys. Rev. Lett.} {\bf 86}, 795--798 (2001).

\bibitem{mul} J. Mulet, C.R. Mirasso, T. and I. Fischer
``Synchronization scenario of two distant mutually coupled semiconductor lasers",
{\em J. Opt. B: Quantum Semiclass. Opt.} {\bf 6}, 97 (2004). 

\bibitem{tar} G.H.M. Tartwijk and D. Lenstra,
``Semiconductor lasers with optical injection and feedback",
{\em Quantum Semiclass. Opt.} {\bf 7}, 87--143 (1995).

\bibitem{suk} D.W.Sukow, T.Heil, I.Fischer, A. Gavrielides, A. Hohl-AbiChedid and W. Els{\"a}sser,
``Picosecond intensity statistics of semiconductor lasers operating in the low-frequency fluctuation regime",
{\em Phys. Rev. A} {\bf 60}, 667 (1999).

\bibitem{san94} T. Sano,
``Antimode dynamics and chaotic itinerancy in the coherence collapse 
of semiconductor lasers with optical feedback"
{\em Phys. Rev. A} {\bf 50}, 2719.–2726 (1994).
 
\bibitem{bul06} J.M. Buld\'u, T. Heil, I. Fischer, M.C. Torrent and J. Garc\'{\i}a-Ojalvo, 
``Episodic synchronization via dynamical injection", 
{\em Phys. Rev. Lett.} {\bf 96}, 024102 (2006).

\bibitem{fis06} I. Fischer, R. Vicente, J.M. Buld\'u, M. Peil, M.C. Torrent, 
C.R. Mirasso and J. Garc\'{\i}a-Ojalvo, 
"Zero-lag long-range synchronization via dynamical relaying",
{\em Phys. Rev. Lett.}, in press (2006). 

\end{thebibliography}
\end{document}